\documentclass[11pt]{article}
\usepackage{graphicx}
\usepackage[margin=1.25in]{geometry}
\usepackage[usenames,dvipsnames]{color}
\usepackage{url}
\usepackage[colorlinks = true,
            linkcolor = blue,
            urlcolor  = blue,
            citecolor = blue,
            anchorcolor = blue]{hyperref}

\usepackage{todonotes}

\makeatletter
\renewcommand\paragraph{\@startsection{paragraph}{4}{\z@}%
            {-2.5ex\@plus -1ex \@minus -.25ex}%
            {1.25ex \@plus .25ex}%
            {\normalfont\normalsize\bfseries}}
\makeatother

\newcommand\pubnumber{Preprint}
\newcommand\pubdate{March 30, 2022}

\def\Title#1{\begin{center} {\LARGE #1 } \end{center}}
\def\Author#1{\begin{center}{ \sc #1} \end{center}}
\def\Address#1{\begin{center}{ \it #1} \end{center}}

\newcommand\pubblock{\rightline{\begin{tabular}{l} \pubnumber\\
         \pubdate \end{tabular}}}
\newenvironment{Abstract}{\begin{quotation} \begin{center}
                       ABSTRACT
     \end{center}\bigskip  }{\end{quotation}}

\def\beq{\begin{equation}}
\def\eeq#1{\label{#1}\end{equation}}
\def\eeqn{\end{equation}}

\newenvironment{Eqnarray}%
   {\arraycolsep 0.14em\begin{eqnarray}}{\end{eqnarray}}
\def\beqa{\begin{Eqnarray}}
\def\eeqa#1{\label{#1}\end{Eqnarray}}
\def\eeqan{\end{Eqnarray}}

\let\bar=\overbar

\def\lsim{\mathrel{\raise.3ex\hbox{$<$\kern-.75em\lower1ex\hbox{$\sim$}}}}
\def\gsim{\mathrel{\raise.3ex\hbox{$>$\kern-.75em\lower1ex\hbox{$\sim$}}}}

\def\L{{\cal L}}

\def\del{\partial}
\def\Dslash{\not{\hbox{\kern-4pt $D$}}}
\def\dslash{\not{\hbox{\kern-2pt $\del$}}}
\def\pslash{\not{\hbox{\kern-2pt $p$}}}
\def\ETmiss{\not{\hbox{\kern-4pt $E$}}_T}

\def\Dlr{\mathrel{\raise1.5ex\hbox{$\leftrightarrow$\kern-1em\lower1.5ex\hbox{$D$}}}}

\def\MSB{{\bar{M \kern -2pt S}}}
\def\msb{{\bar{\scriptsize M \kern -1pt S}}}

\def\drb{{\bar{\scriptsize D \kern -1pt R}}}

\def\GeV{{\rm GeV}}
\def\TeV{{\rm TeV}}

\newcommand\snowmass{\begin{center}\rule[-0.2in]{\hsize}{0.01in}\\\rule{\hsize}{0.01in}\\
\vskip 0.1in Submitted to the  Proceedings of the US Community Study\\ 
on the Future of Particle Physics (Snowmass 2021)\\ 
\rule{\hsize}{0.01in}\\\rule[+0.2in]{\hsize}{0.01in} \end{center}}

\begin{document}

\pubblock

\Title{Particle Flow Calorimetry}

\bigskip 

\Author{Randal Ruchti$^1$, Katja Kr\"uger$^2$}
\Address{$^1$Department of Physics, University of Notre Dame, IN 46556, USA}
\Address{$^2$Deutsches Elektronen-Synchrotron DESY, Notkestr. 85, 22607 Hamburg, Germany}

\medskip

\begin{Abstract}
The motivation for PF calorimetry is to experimentally measure the energy of hadron jets with excellent resolution.  In particle flow designs,  $\sigma/E < 5\%$ should be possible for a range of jet energies from $\sim 50\,\GeV$ to $\sim 250\,\GeV$, important particularly for experiments at electron-positron colliders (ILC, CLIC, FCCee, CEPC).  The high granularity, which is essential for PF calorimetry, can also be very beneficial for removal of background from pile-up on an event-by-event basis making such calorimeters an attractive approach for hadron collider experiments, for example the HGCAL under construction for CMS at the CERN HL-LHC.
\end{Abstract}

\snowmass
 
\clearpage

\section{Introduction}   
\label{section:intro}

In the search for new physics at particle colliders, the understanding of jets and jet reconstruction have been important tools for discovery.  Jet reconstruction has generally been the domain of calorimetry, where particle energies are measured in an ensemble of instruments with separated functionality: electromagnetic calorimeters for measurement of electron, positron and photon energies, and hadron calorimeters for the measurement of charged and neutral hadron energies.   Because of sheer size and volume, the hadron calorimetry designs involved coarse measurement elements, at the expense of being able to measure hadron energies precisely. Nevertheless these calorimeters provided exellent hermeticity, a functionality important for triggering and for the measurement of missing transverse energy.

In recent years, there has been paradigm shift in design favoring high granularity Particle Flow calorimetry, driven in part by the scientific objectives of Higgs factories such as the International Linear Collider, where a premium has been placed on clean separation of the hadronic decays of vector bosons (W, Z) and for precision study of Higgs physics. Such calorimeters are designed to serve as effective extensions of the tracking system with granularity sufficient to associate energy depositions of individual incident charged particles whose momenta have been measured independently in tracking, revealing spatially identifiable energy clusters created by neutral hadrons.  By making the appropriate associations of this new information, jet resolutions of $\sigma_E/E < 30\%/\sqrt(E)$ have been demonstrated to be achievable. 

\section{Motivations for Particle Flow Calorimetry}
\label{section:motivation}

While there are commonalities in the choice of Particle Flow Calorimetry designs across a variety of experiments, the motivations behind the choices vary.  In common are a sampling-layered structure with very high granularity of readout channels, so that energy deposits recorded in the calorimetry can be associated correctly with charged particles whose momenta are recorded in tracking detectors. EM objects (e, $\gamma$) are identified and registered primarily through energy deposits in the initial layers and charged hadrons registered over a significantly greater range of layers in depth.  Lastly, unassociated energy deposits are then ascribed to neutral hadrons.  In lepton colliders (ILC, CLIC), the high granularity is used to emphasize separation among the different objects and avoid confusion created by overlapping information.   At hadron colliders (HL-LHC/CMS), a major concern involves event pileup created by the very high luminosity and the need to ameliorate such effects.   In both cases, extracting the physics performance from the wealth of information depends upon complex pattern recognition analyses that ultimately yield a list of reconstructed objects including particle type, energy and momentum. 

\subsection{Scientific Motivation}

The scientific goals of $e^+e^-$ colliders (linear or circular) are directed toward reconstruction and precision measurement of specific physics final states - for example, W, Z and Higgs - in a relatively clean initial state environment.   Variously for hadron colliders, emphasis is directed toward searches for and discovery of new physics, generally in the face of consequential event pileup and radiation fields.  Interestingly, both can benefit strongly from a new holistic view of tracking and calorimetry design which are the domain of particle flow techniques.

\subsubsection{Vector Bosons and Jet Reconstruction}
For linear collider experiments, a performance benchmark is to distinguish W, Z based on their (dominant) hadronic decays. To do so requires a hadronic mass resolution comparable to the gauge boson widths, $\sigma_{\rm m}/{\rm m} \approx 2.7\%$.  That this is achievable has been demonstrated for the ILD detector's  hadron calorimeter using software compensation~\cite{Tran:2017tgr}, based on simulation validated with CALICE testbeam data.

For hadron collider experiments, VB Fusion and VB scattering processes, top quark physics, Higgs physics and searches for evidence of BSM physics will continue to be prominent areas of investigation.  The granularity of the CMS HCAL should allow the ability to distinguish quark and gluon jets, important for better understanding of VBF events~\cite{CERN-LHCC-2017-023}.

\subsubsection{Higgs Production and Event Reconstruction}
At linear colliders, precision measurement of the Higgs boson and its various decay channels are major priorities.  The predominant production mechanisms of Higgs bosons in the Standard Model (SM) are Higgsstrahlung at collision energies below 450 GeV and WW fusion at higher energies~\cite{ILCHIGGSWP, ILCHIGGSWP2021}. In the Higgsstrahlung case, reconstruction of the Z is the crucial piece.  Since the kinematics are constrained, the recoil mass allows "identifying" the Higgs without selecting a decay channel or reconstructing it directly.  Should the Z decay hadronically, the challenge is to define a Z selection that has the same efficiency for all Higgs decays, in the presence of an unknown number of additional jets.  In the case of WW fusion, the challenge is reconstruction of Higgs pairs with missing energy.

At the HL-LHC, an important objective is the measurement of the Higgs self-coupling. Given the expected rarity of this process, being able to combine as many possible decay channels for the Higgs decay will be of significant benefit.  Additionally,  hadron colliders are intrinsically discovery machines due to their energy reach, and several BSM theories suggest that there may be more Higgs-like particles than only the one in the SM.  Discovery will  place strong demands on excellent jet reconstruction, jet resolution and angular measurement.

\subsubsection{Tau Reconstruction}

Calorimetry plays a role in tau lepton detection and reconstruction primarily through its hadronic decays, which consist of charged and neutral pions and an associated (and undetected) neutrino.  The granularity of PF calorimetry is beneficial to the reconstruction of these narrow jets, including identification and measurement of gamma energies, and, with timing  timing capability, to assist with pileup background rejection~\cite{CERN-LHCC-2017-023}.

\subsubsection{Long Lived Particles}

The search for long lived particles can be facilitated with PF calorimeters that are subsystems of multipurpose collider detectors such as those at hadron or lepton machines, or in specialized designs (for example for forward physics).  The benefits of PF design include energy and timing measurement and pointing, the ability to identify the source of the energy with a vertex location that is significantly displaced from the source of primary production.  The actual effectiveness of any such approach will depend in part upon the rest mass and lifetime of such hypothetical particles and the effective available decay length in a given experiment.

\subsection{Technical Motivation and Challenges}
\subsubsection{Channel Count and Calibration}
Since the strength of PF calorimetry is its effective extension of particle tracking into the energy measurement, the substantial channel count requires special attention to calibration.  For the hadron sections, calibration will be effected in part by detection of mip signals (muons and charged hadrons) and by electrons and positron detection in the EM sections. Critical issues are aging due to radiation and thermal/mechanical issues affecting sensors and associated electronics.

\subsubsection{Event Timing, Pileup Subtraction and Vertex Location}
In  $e^+e^-$ linear colliders, beam collisions set the event timing, but timing capability can also serve as an aid in rejecting beam-gas interactions.  In hadron colliders, the large number of collisions per beam crossing (140-200 at HL-LHC) leads to significant event pileup. The ability to measure timing of particles as well as their energies becomes an important tool for event reconstruction, association of particles to primary or secondary vertices and for pileup background suppression.

\section{Overview of PF Calorimeters currently under development or construction}
\label{section:developement}

Particle Flow calorimeters are under development for future collider applications as well for upgrades of existing detectors.  And numerous detector technologies are under development to extend the reach of such instrumentation. Several of these important efforts have been submitted and described by the proponents in Letters of Interest to the Snowmass2021 Community Summer Study and are highlighted here.

\subsection{Detector Upgrades/Concepts}

\subsubsection{CALICE}
\label{subsub:CALICE}
The CALICE collaboration is a dedicated detector R\&D collaboration focusing on the development of highly granular calorimeters optimized for the application of particle flow algorithms. Originally founded for calorimeter development for the International Linear Collider (ILC), the scope has been widened to also integrate applications of highly granular calorimeters in other environments. 
CALICE aims to provide a framework for the R\&D activities, ensuring common standards for beam tests and data analysis, as well as organizing combined beam tests of full calorimeter systems consisting of ECAL, HCAL and tail catcher/muon tracker. For the ECAL, small scintillator strips and silicon sensors are investigated as active material. For the HCAL, scintillator tiles and gaseous detectors (RPCs, GEMs and MicroMegas) are studied. For the ECAL, tungsten is chosen as absorber due to its very small Moliere radius. For the HCAL, the main absorber option is stainless steel for its structural strength and machinability, while tungsten is considered as alternative for very high center-of-mass interactions to limit the size of the calorimeter while keeping sufficient thickness in interaction lengths. 

Since the typical energy depositions in the cells are small, various readout concepts have been developed: in addition to a “traditional” analog readout with detailed information for each cell energy, a digital approach, recording only if the energy crossed a fixed threshold, and a semi-digital approach, recording hits with three thresholds, have been developed and tested. For several of the concepts, large prototypes have been built. The first generation of prototypes focused on demonstrating the performance reachable with the corresponding concept. The second generation has focused on the technical implementation of the concept scalable to a full collider detector, which requires solutions adapted to a specific application and environment. The requirement of scalability leads to the front-end electronics being integrated into the active layers. One example of an adaptation to a specific environment is power pulsing, which allows the building of  calorimeters with minimal-to-no active cooling in the active layers at linear $e^+e^-$ colliders.   By adapting the front-end electronics to the beam structure of the machine, the electronics is only switched on during the bunch train ($1\,{\rm ms}$ at ILC), while it is off to save power in the long gaps between bunch trains ($199\,{\rm ms}$). 

The comparison of the test beam data with simulations forms an integral part of the CALICE mission, both to provide a detailed realistic and validated modeling of the concepts and to provide feedback to the simulation model developers. 
Recently, dedicated efforts to measure the time development of showers and to improve the resolution of the hit time measurements have been initiated. 
Details of the individual technologies will be discussed in the next section.

\subsubsection{ILD}
\label{subsub:ILD}
The ILD is a multipurpose detector and one of the two validated detector designs for the future ILC. The particle flow paradigm was adopted as its guiding principle since the start and the design has developed around it. 

ILD has chosen a sampling calorimeter equipped with silicon diodes as one option for the electromagnetic calorimeter. Diodes with pads of about $5 \times 5\,{\rm mm}^2$ are used, to sample a shower up to $30$ times in the electromagnetic section. A self-sustaining structure holding tungsten plates with Carbon-Fiber-Reinforced-Polymer (CFRP) supports the detector elements while minimizing non-instrumented spaces. As an alternative to the silicon based system, sensitive layers made from thin scintillator strips are also investigated. Orienting the strips perpendicular to each other has the potential to realize an effective cell size of $5 \times 5\,{\rm mm}^2$, with the number of read-out channels reduced by an order of magnitude compared to the all silicon case.

For the hadronic part of the calorimeter of the ILD detector, two technologies are studied, based on either silicon photomultiplier (SiPM) on scintillator tile technology or resistive plate chambers. The SiPM-on-tile option has a moderate granularity, with $3 \times 3\,{\rm cm}^2$ tiles, and provides an analogue readout of the signal in each tile (AHCAL). The RPC technology has a higher granularity, of $1 \times 1\,{\rm cm}^2$, but provides only 2-bit amplitude information (SDHCAL). 

The conceptual and technological development of this particle flow calorimeter have been largely done by the CALICE collaboration.

\subsubsection{SiD}
\label{subsub:SiD}
The SiD Detector (Snowmass2021 LoI~\cite{LoI:SiD})  is the other of the two validated detector designs for the future ILC. SiD features a compact, cost-constrained design for precision Higgs and other measurements, and sensitivity to a wide range of possible new phenomena. A robust silicon vertex and tracking system, combined with a $5\,{\rm T}$ central solenoidal field, provides excellent momentum resolution. The highly granular calorimeter system is optimized for Particle Flow  to achieve excellent jet energy resolution over a wide range of energies. 

The SiD ECAL design employs thirty silicon longitudinal layers, the first twenty each with a $2.50\,{\rm mm}$ tungsten alloy absorber thickness and $1.25\,{\rm mm}$ readout gaps, and the last ten with $5.00\,{\rm mm}$ tungsten alloy. The total depth is $26$ radiation lengths, providing good containment of electromagnetic showers. The baseline design employs tiled, large, commercially produced silicon sensors. The sensors are segmented into pixels that are individually read out over the full range of charge depositions. The complete electronics for the pixels is contained in a single chip, the KPiX ASIC, which is bump bonded to the wafer. The low beam-crossing duty cycle allows reducing the heat load using power pulsing, thus allowing passive thermal management within the ECAL modules. Results have been obtained from an initial 9-layer prototype and further R\&D is foreseen on a full-depth stack.

The SiD HCAL has a depth of $4.5$ nuclear interaction lengths, consisting of alternating steel plates and active layers. The baseline choice for the active layers is scintillator tiles read out via silicon photomultipliers. For this approach SiD is closely following the developments within the CALICE collaboration. An initial, conceptual mechanical design has been developed, and further work is needed on all aspects of the design.

\subsubsection{CLICdet and CLD}
\label{subsub:CLICdet}
CLIC is a proposed linear electron-positron collider with center-of-mass energies up to $3\,\TeV$, while the proposed circular electron-positron collider FCC-ee reaches only lower energies, but offers unprecedented luminosities at the Z mass. Both can operate at energies close to the highest Higgsstrahlung and top-antitop production cross sections, and are proposed for construction in Europe (at CERN). The CLICdp collaboration has developed the CLICdet detector concept and is actively engaged in the corresponding detector technology development. In recent years, members of the CLICdp collaboration and the FCC collaboration have adapted the CLIC detector concept for operation at FCC-ee, resulting in the CLD concept, optimized in view of the FCC-ee Higgs and top physics energy stages.

Both CLICdet and CLD (Snowmass2021 LoI~\cite{LoI:CLICdet}) are general purpose $e^+e^-$ detector concepts. Their design was driven by jet energy resolution (through PFA), together with high-performance tracking in view of precision measurements of single particles and jet flavor identification. The reconstruction of individual particles in a dense environment through PFA is also essential for the suppression of particles from beam-induced background in $e^+e^-$ colliders, which is especially relevant in the forward direction. For the ECAL, a sampling calorimeter with tungsten absorbers and silicon active layers of $5 \times 5\,{\rm mm}^2$ cell size is foreseen, comprising $40$ layers. For the HCAL, a sampling calorimeter with steel absorbers and scintillator active layers with SiPM readout is planned, with a cell size of $3 \times 3\,{\rm cm}^2$. While the CLICdet HCAL has $60$ layers to ensure a good containment of very high energetic jets, the CLD HCAL has only $44$ layers. In the domain of calorimetry, the technology development is carried out in the framework of the CALICE and FCAL collaborations. A notable distinction to these detector developments compared to those for ILC applications is the nature of the beam structure, which does not allow for power pulsing (FCCee) or requires adaptations to it (CLIC).

\subsubsection{CEPC}
\label{subsub:CEPC}
The Circular Electron Positron Collider (CEPC) is proposed for construction in China as a Higgs factory to study the Higgs boson with an unprecedented precision. It also serves as a Z-factory when operating at the Z pole. The CEPC baseline detector concept includes a PFA calorimetry system (Snowmass2021 LoI~\cite{LoI:CEPC}). The design implements concepts developed within the CALICE collaboration. There are still various detector technology options that need to be further explored to address challenges from the stringent performance requirements for the CEPC detector design. 
\begin{itemize}
  
\item{ECAL:} A highly granular calorimeter with a scintillator-tungsten sandwich structure is one option for the PFA ECAL in the CEPC baseline detector conceptual design. The design largely follows that of the ILD Sci-W ECAL, and realizes a very high effective granularity with less channels by using scintillator strips arranged in a way that strips in adjacent layers are orthogonal to each other. 

\item{HCAL:} The Sci-Fe HCAL concept is adopted as one PFA HCAL option. There are significant similarities between the Sci-Fe HCAL and Sci-W ECAL detector options in that both use the technology of scintillator directly coupled to SiPM. A major difference is the different scintillator cell size and geometry. While the Sci-W ECAL uses small scintillator strips, the Sci-Fe HCAL uses large scintillator tiles. 

\item{Active cooling system:} Due to the high granularity design of PFA calorimeters, the number of readout channels of the calorimetry system is extremely large. Because CEPC operates in continuous mode, the power dissipation produced by the embedded readout system could reach about 100kW. It is therefore crucial to design an active cooling system to bring the heat out of the CEPC PFA calorimeters. 
\end{itemize}

\subsubsection{CMS HGCAL}
\label{subsub:HGCAL}
The Phase 2 upgrades of the Compact Muon Solenoid (CMS) detector will enable the detector to function in the environment of the High Luminosity Large Hadron Collider (HL-LHC). Slated to begin operations in 2027, the HL-LHC will have an instantaneous luminosity of $\L = 5 - 7.5 \times 10^{34}\,{\rm cm}^{-2}{\rm s}^{-1}$, a projected total integrated luminosity of $3 - 4.5\,{\rm ab}^{-1}$, and 140 - 200 pileup interactions per bunch crossing. These benchmarks significantly exceed those of the LHC, for which the current CMS detector was designed. The Phase 2 upgrade program largely encompasses improvements to the data acquisition (DAQ) system to increase the Level-1 trigger accept rate from 100 to 750 kHz and replacements of aging front end materials with new ones expected to withstand the HL-LHC total ionizing dose and 1-MeV-equivalent neutron fluence in the CMS interaction region. 

\begin{figure}[htb]
\begin{center}
\includegraphics[width=0.50\hsize]{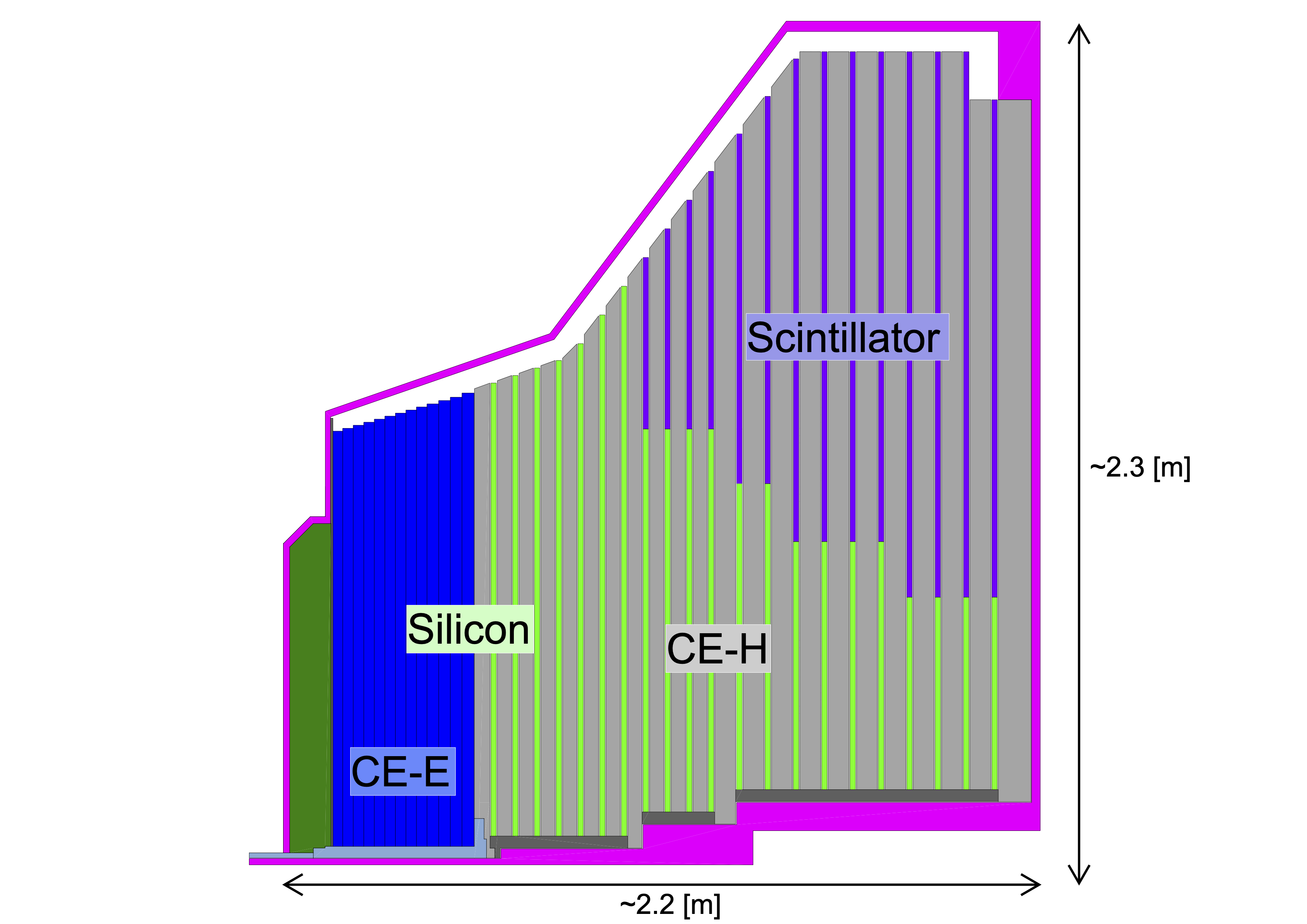}
\hfill
\end{center}
\caption{layer structure of the HGCAL}
\label{fig:HGCAL}
\end{figure}

Central to the Phase 2 upgrades is the replacement of the current endcap ($1.5 \leq |\eta| \leq 3.0$) calorimeter, consisting of a total absorption scintillating crystal electromagnetic section followed by a brass and plastic scintillator sampling hadronic section, with the High Granularity Calorimeter (HGCAL) (Snowmass2021 LoI~\cite{LoI:HGCAL}), which is a 6M channel particle flow sampling calorimeter. The HGCAL is a $47$-layer sampling calorimeter with 26 layers constituting a front electromagnetic section (CE-E) of 27.7 radiation lengths ($X_0$) followed by $21$ layers composing a rear hadronic section (CE-H).  The total calorimeter depth is $9.97$ hadronic interaction lengths ($\lambda_I$). Copper, tungsten, and lead are used as absorber materials in the CE-E, while stainless steel is used in the CE-H. Detector planes equipped with either silicon sensors or scintillator tiles with SiPM readout are used. The silicon sensors will account for $\sim 55\%$  of the total active area, which is $1100\,{\rm m}^2$. The silicon is divided into $\sim 6$ million channels with an area of either $\sim 0.5$ or $\sim 1\,{\rm cm}^2$. The CE-H design foresees $\sim $400,000 scintillator+SiPM channels. The longitudinal cross section of one half of one endcap of the HGCAL design is shown in Figure~\ref{fig:HGCAL}.

After the final phases of sensor prototyping, major construction is planned to begin in 2022. Drawing on years of R\&D pioneered by the linear $e^+e^-$ collider community, the HGCAL will be the largest PF calorimeter ever built, and the first to operate at a hadron collider. Throughout the next decade, the lessons of the HGCAL project -- how to construct high quality, highly segmented calorimeters on a budget; and the utility of PF design at high luminosity hadron colliders especially in pileup rejection and radiation tolerance -- will be instrumental in efforts to adapt the PF technology to ambitious new experiments. The HGCAL Collaboration supports continued investment in PF calorimetry, both in direct support of HL-LHC construction and operations and at the R\&D level. Active communication continues between the $e^+e^-$ and hadron collider calorimetry communities, to the benefit of both.

\subsection{Technologies}

\subsubsection{Scintillator Strip ECAL}
\label{subsub:ScECAL}
A calorimeter based on scintillator strips with SiPM readout (Sci-ECAL) (Snowmass2021 LoI~\cite{LoI:ScintStrip}) is one of the promising technology options for a highly granular ECAL being developed within the CALICE collaboration. The detection layers with scintillator strips ($45 \times 5 \times 2\,{\rm mm}^3$ each) coupled to SiPMs are stacked alternately in an orthogonal orientation to achieve an effective transverse segmentation of $5 \times 5\,{\rm mm}^2$, allowing to significantly reduce the number of readout channels. The same performance as the real segmentation of $5 \times 5\,{\rm mm}^2$ is achievable with an appropriate reconstruction algorithm. The detection layers are interleaved by tungsten absorber layers. The Sci-ECAL with the compact and cost-effective design is proposed for ILC-ILD and CEPC-ECAL.

Important progress has recently been made on the SiPMs for Sci-ECAL. MPPCs with a smaller pixel pitch of $10$ or $15\,{\rm \mu m}$ have been developed, which can provide a larger dynamic range required for Sci-ECAL. Further improvements have been made for the most recent small-pixel MPPCs, including reduced optical cross-talk by a trench structure between pixels, lower dark noise and higher photon detection efficiency, which have been confirmed by lab tests. In earlier studies, the SiPM was attached to the side edge of the strip. New designs of the SiPM readout at the bottom side of the strip are being developed for more uniform response and a better compatibility with future large-scale production.

A full-size Sci-W ECAL prototype with readout electronics embedded in the detector has been constructed as a joint effort of the CEPC calorimeter group and some ILC-ILD detector R\&D groups from Japan. The key parameters for this technological Sci-W ECAL prototype are $22 \times 22\,{\rm cm}^2$ for transverse area, 24 $X_0$ for thickness, and $5 \times 5\,{\rm mm}^2$ for effective cell size. The scintillator strips are read out using SiPMs with a small pixel size of $10$ or $15\,{\rm \mu m}$  to provide a large dynamic range for measurement of high energy EM showers. The prototype has been fully assembled and integrated with readout electronics and DAQ. A LED-based SiPM monitoring and calibration system has also been developed and integrated into the readout electronics. The prototype is now being commissioned with cosmic rays and is ready for test beams in the coming years.

The remaining challenges include the development of a detector assembly system for the large-scale production, testing the power pulsing operation of the integrated electronics for the ILC Sci-ECAL and optimizing the cooling system for the integrated electronics for the CEPC Sci-ECAL, where a continuous operation is required unlike the ILC Sci-ECAL.

\subsubsection{Silicon Tungsten ECAL}
A highly granular ECAL based on silicon sensors and tungsten absorber combines the advantages of a very compact design and a possibly radiation hard technology. Silicon-tungsten ECALs are under development for detectors at $e^+e^-$ colliders as well as hadron colliders. The CALICE collaboration is developing one of the concepts (Snowmass2021 LoI~\cite{LoI:SiW-ECAL}), described in more detail here. For other developments see section~\ref{subsub:SiD} (SiD) and~\ref{subsub:HGCAL} (CMS HGCAL).

After the realization of a physics prototype (2005–2011) as a proof of concept, CALICE has constructed a technological prototype of a highly granular silicon tungsten electromagnetic calorimeter (SiW-ECAL), taking into account the many constraints of a large scale detector, using ILD as a baseline. The most critical part is the so-called Active Sensor Unit (ASU), the base element of detection, including a PCB, glued Si sensors, and ASICs. A calorimeter stack with around $20$ layers, each composed of a single ASU, of size $18 \times 18 \times 0.5\,{\rm cm}^3$ with a total depth of about one interaction length is available. The ASUs are subdivided into 1024 readout cells of around $5 \times 5\,{\rm mm}^2$. Their embedded front-end readout electronics records the energy deposit and the time at which a cell is hit, with $12$-bit resolution on a large dynamic range ($1-3000$ MIPs) and a time precision of $\sim 1.5\,{\rm ns}$ for a MIP.

In the coming years, beam tests in conditions compatible with the ILC (power pulsing) will be used to fully characterize the prototype in terms of linearity uniformity, noise, and cross talk, for MIPs, and electromagnetic and hadronic showers. Detailed characterization of the spatial shape of electromagnetic and hadronic cascades in tungsten, will be complemented by the timing information. Combined beam tests are foreseen with already existing prototypes of hadronic calorimeters. With an overall depth of about $6$ interaction lengths, starting with the much higher segmentation of the ECAL, where $55\%$ of hadrons interact, these setups are an important validation of PFA performance with realistic devices.

Future R\&D for the SiW-ECAL will address several aspects: (1) System issues to integrate ASUs into long cassettes of up-to twelve chained ASUs; (2) Large sensors to profit from the use of 8" wafers pioneered by the CMS~HGCAL; (3) Thin design by moving from BGA packaged ASICs (overall ASU thickness of $\sim 3\,{\rm mm}$) to ultra-thin ASU, $1.2\,{\rm  mm}$ thick, where the ASICs are mounted in recessed cavities; and (4) Reliability to ensure that all components of the detector (connectors, flat capacitors, ...) work reliably in an experiment for more than twenty years, possibly operated in a pulsed mode in a strong magnetic field.

\subsubsection{Digital Silicon ECAL}
Within the ALICE collaboration there is a proposal to develop and build a forward calorimeter, FoCal, including a SiW electromagnetic part with several layers of extremely high granularity. In the context of the R\&D for this calorimeter, and as part of the general R\&D effort in the CALICE collaboration, a SiW calorimeter based on monolithic active pixel sensors with a pixel pitch of $30\,{\rm \mu m}$  is being developed (Snowmass2021 LoI~\cite{LoI:DECAL}). A first prototype with $\sim 40$ million pixels has been constructed using MIMOSA sensors and has successfully been tested in beam, showing good energy linearity and resolution and extremely high position resolution and two-shower separation capabilities. This provided a proof of principle of digital pixel calorimetry only, because the MIMOSA sensors are too slow to use in a collider experiment.

Very recently a second full pixel prototype has been constructed using ALPIDE sensors, which have been developed at CERN for the ALICE ITS upgrade. These sensors are fast enough to be used in the ALICE FoCal upgrade, and first tests show excellent calorimetric performance. It is planned to go further with this development to explore the possibilities of still faster sensors for calorimetric purposes that could be used in a standard modern collider experiment. This would be of interest for detectors at future hadron and electron-positron colliders, and also for EIC. Further development could be based on a next generation of the ALPIDE technology, but also other pixel sensors could be suitable.
The detector concept in development for FoCal has been shown to have excellent performance for direct photon measurement, requiring in particular the separation of close-by EM showers. The performance for other applications of the detector have not really been explored so far. In parallel to the technological R\&D, the possible performance of such a technology for jet measurements will be studied, in particular with suitably adapted particle flow algorithms.
A very similar detector concept using the same technology is the basis for a development of a proton CT scanner for medical applications.

\subsubsection{Gaseous HCALs}
To cover the large areas of the detection layers in a highly granular hadron calorimeter, gas-based detectors like Resistive Plate Chambers (RPCs), gas electron multipliers (GEMs) or Micromegas provide a cost-effective solution. For gaseous detectors operated in avalanche mode, the measured charge is only weakly correlated with the primary deposited energy in a given cell, such that the detailed readout of individual hit charges is not expected to improve the hadron energy resolution. Therefore, dedicated readout concepts with digital or semi-digital hit information have been developed. 

\paragraph{Resistive Plate Chambers} 
Within the CALICE collaboration two large prototypes with RPCs as active media have been built.
The CALICE Digital Hadron Calorimeter (DHCAL) (Snowmass2021 LoI~\cite{LoI:DHCAL}) uses RPCs as active media and has a 1-bit resolution (digital) readout of $1 \times 1\,{\rm cm}^2$ pads. A prototype has been built and tested with steel and tungsten absorber structures, as well as with no absorber structure.
A single layer of the DHCAL prototype measures roughly $1 \times 1\,{\rm m}^2$ and consists of $96 \times 96$ pads, with the readout electronics embedded in the active layer. In addition to the absorber plates, each layer of RPCs was contained in a cassette with a $2\,{\rm mm}$ thick Copper front plate and a $2\,{\rm mm}$ thick steel back plate. 
In order to obtain a unique dataset of electromagnetic and hadronic interactions with unprecedented spatial resolution, the DHCAL went through a broad test beam program. For test with steel absorber, up to $52$ layers were installed. The calorimeter consisted of a $38$-layer main stack with $1.75\,{\rm cm}$ thick steel absorber plates and a $14$-layer tail catcher. The tungsten DHCAL consisted of $54$ active layers, where the first $39$ of these layers were inserted into a tungsten absorber structure featuring $1\,{\rm cm}$ thick tungsten plates. The remaining $15$ layers were inserted into a steel tail catcher. The total thickness of the $54$-layer W-DHCAL (main stack and tail catcher) corresponded to approximately $183$ radiation lengths or $11.1$ nuclear interaction lengths. In special tests, $50$ active layers of the DHCAL were exposed to low energy particle beams, without being interleaved by absorber plates. The high spatial segmentation of the imaging calorimeters allows the application of corrections to the measured number of hits which might result in an improved linearity of the response and energy resolution. The DHCAL data can be corrected for leakage using the detailed shower profiles, and the response can be linearized by assigning different weights to hits based on their densities.

The CALICE Semi-Digital Hadron Calorimeter (SDHCAL) also uses RPCs as active media, but has a 2-bit resolution (semi-digital) readout. A large prototype of $1.3{\rm m}^3$ made of 48 units was designed and built. Each unit is built of an active layer made of $1{\rm m}^2$ glass RPC detector placed inside a cassette whose walls are made of stainless steel. The cassette contains also the electronics used to read out the RPC detector, which support operation in power pulsing mode. The lateral granularity of the active layer is provided by the electronics pick-up pads of $1\,{\rm cm}^2$ each. The cassettes are inserted into a self-supporting mechanical structure built also of stainless steel plates which, with the cassettes walls, play the role of the absorber.  A dedicated acquisition system was developed to deal with the output of more than 440,000 electronics channels in both trigger and triggerless modes. Since 2012, the prototype has been subject to a large scale test beam campaign. The numbers of hits above the three thresholds can be used to reach a linear response (within a few percent) and a good energy resolution for a large range of hadron shower energies. 

The CALICE collaboration has also performed tests of GEMs and Micromegas in smaller prototypes, which have shown the general suitability of these technologies for application in a digital or semi-digital hadron calorimeter.

\paragraph{Advanced GEM Detectors}
Micro-Pattern Gaseous Detectors (MPGDs) (Snowmass2021 LoI~\cite{LoI:AdvancedGEM}) are radiation hard detectors, able to cope with rates of ${\rm MHz/cm}^2$, exhibiting good spatial resolution ($\leq 50\,{\rm \mu m}$) and good time resolution of $5-10\,{\rm ns}$. They have the potential of economically covering large areas. Gas Electron Multiplier (GEM) detectors are the most consolidated technology inside the MPGD family, and in the last decade they have been considered as tracking devices for Muon Systems in LHC experiments. Additionally, several technological solutions and optimization of the operation parameters have been achieved in order to make the detector suitable for operation at HL-LHC. 

The challenge is now open for future colliders considering the high particle rates, discharge probabilities and accumulated doses expected at Future Colliders. Time resolutions of $30-100\,{\rm ps}$ per track are required to distinguish pile-up collisions and high background conditions, while sub-ns time resolutions are required for experiments at future colliders that will work at $200\,{\rm MHz}$ bunch crossing frequency. Studies are planned that will aim at achieving major advances in the development of a new generation of MPGDs for several applications in HEP experiments at future colliders, such as a pre-shower for an electromagnetic calorimeter, active readout layers in sampling calorimeters, a muon tracking or tagging detector at FCC-ee/hh and Muon Collider or as the readout layer of a Time Projection Chamber. 

The main objectives of the R\&D are: (1) stable and efficient operation with incident particle flows up to $\sim 100\,{\rm MHz/cm}^2$; (2)  radiation hardness up to integrated charges of hundreds of ${\rm C/cm}^2$; (3) very high granularity readout with dedicated electronics to operate efficiently at high fluxes; (4) high timing resolution; and (5) manufacturing on an industrial scale.
The main technological solution explored to optimize the operation of MPGD detectors in terms of speed and uniformity of response, damping and eliminating gas discharges, spatial resolution and gain, under particle flows expected at future accelerators, is the development of GEM foils with resistive electrodes.
Industrial partners are already involved in the development of the new technology with two advantages: being at the forefront in the mass production phase of the detectors to be installed at future colliders and taking advantage of any spin-offs of these technologies in application areas outside HEP such as X-ray and neutron imaging systems in the industrial field, hadron therapy in the medical field, muon tomography in geology and archaeology, and for national security systems.

\subsubsection{Scintillator HCAL}

The CALICE collaboration has pioneered the development of the SiPM-on-Tile technology for highly granular calorimeters (Snowmass2021 LoI~\cite{LoI:AHCAL}). The construction and operation of a Physics Prototype of the Analog Hadron Calorimeter (AHCAL) established the performance, in terms of energy and topology reconstruction, of the approach in principle, as well as the adequacy of detailed simulations. An AHCAL Technological Prototype, which demonstrates the viability of engineering solutions for a large collider detector, has been built and successfully tested in beams at CERN in 2018. The prototype consists of a steel absorber structure and active layers of small scintillator tiles that are individually read out by directly coupled SiPMs. Each layer has an active area of $72 \times 72\,{\rm cm}^{2}$ and a tile size of $3 \times 3\,{\rm cm}^{2}$. With $38$ active layers, the prototype has nearly $22,000$ readout channels, and its total thickness amounts to $4.4$ nuclear interaction lengths. The dedicated readout electronics provide time stamping of each hit with an expected resolution of about~$1\,{\rm ns}$.
The design has been guided by the ILD detector concept for the ILC, but variations are being proposed for SiD at the ILC, CLICdet at CLIC and CLD at FCCee as well. Moreover, it has inspired the design chosen for the scintillator section of the upgrade of the CMS endcap calorimeter (HGCAL) for the HL-LHC.

Further beam measurements are planned with the prototype to fully exploit the ns timing capabilities of the technological prototype readout electronics, to study hadron showers in an existing tungsten absorber structure and to study the performance for energy and topology reconstruction in a realistic configuration with a combined ECAL and HCAL system. In addition, the AHCAL prototype can be used as a test bed for new types of scintillators, SiPMs, front end electronics and back-end interfaces, providing a realistic environment and, by virtue of the reconstruction of the shower start location, enabling the test of components in different stages of the shower evolution. Timing layer prototypes with tens of ps resolution could be integrated in the prototype to study their interplay with the rest of the detector showing more moderate resolution.

To validate the AHCAL technology option for its application to the CEPC experiment, a large-scale technological prototype is being developed by the CEPC calorimeter group. It consists of 40 longitudinal sampling layers with a transverse plane size of $\rm 72\times72~cm^2$ and a total depth of around $4.8\lambda_{I}$. There are 12,960 SiPM-on-Tile readout channels and the total weight is around 4.5 tons. The dimension of a scintillator tile is $\rm{40\times40\times3~mm^{3}}$ and the cavity inside the tile for coupling is optimized for the NDL-SiPM package to improve the response uniformity. The prototype design optimization studies have been done for the CEPC based on the Arbor PFA, including the cell size, the number of sampling layers, total thickness of the absorber, the thickness of scintillator tiles, etc. The "CEPC-AHCAL" prototype has the major updates: (1) tile size is optimized by PFA studies to $\rm 40\times40\times3~mm^3$ from $\rm 30\times30\times3~mm^3$, with coarser granularity but similar performance expected; (2) to cover the same transverse size of $\rm 72\times72~cm^2$, the dimensions of HBU PCBs has a different shape and considerably larger i.e., $\rm 24\times 72~cm^3$ instead of $\rm 36\times 36~cm^2$; (3) besides the ones from HPK, SiPMs from another vendor "NDL" with a high pixel density and a large sensitive area have been instrumented to achieve a better signal-to-noise ratio and a wider dynamic range; (4) the CEPC-AHCAL DAQ system has been newly designed and optimized to be capable to read out events with a much higher rate, up to around 3kHz (finally limited by the SPIROC chip). A fully integrated system dedicated to SiPM calibration and monitoring has been developed. The QA/QC system for scintillator tiles and SiPMs have been established and successfully tested all needed components. The construction of the technological prototype is expected to be completed in 2022. After finishing the commissioning with cosmic muons, the prototype will then be tested with high energy hadron beams at CERN in late 2022 and/or early 2023. 

\subsubsection{Highly Granular Crystals}
A highly granular crystal electromagnetic calorimeter (ECAL) (Snowmass2021 LoI~\cite{LoI:HGCrystal}) is newly proposed in the context of the PFA-oriented detector design for future lepton colliders, with an aim to achieve an optimal electromagnetic energy resolution (around $3\%/\sqrt{E(GeV)} \oplus 1\%$) using the homogeneous design with scintillation crystals while maintaining the PFA capability for precision measurements of jets with finely segmented crystal cells. The electron momentum resolution can be improved with Bremsstrahlung corrections to refine Higgs recoil mass measurements. With the capability to trigger on single photons, crystal calorimetry can also play an essential role in the rich flavour physics programs as well as searches for new physics. Dedicated readout schemes to determine Cherenkov and scintillation photons in crystals with the information of the wavelengths or timing are expected to significantly improve the energy resolution of hadrons and jets.

To strengthen the joint efforts, software developments and design optimizations are indispensable. Small-scale hardware prototypes and measurements can also provide solid inputs to validate the simulations. An inexhaustible list of key issues and critical R\&D tasks have been initiated: (1) Develop reliable software tools to validate the crystal calorimeter and reconstruction algorithms optimized for different layouts; (2) Study the electromagnetic showers with realistic detector layouts and calibration schemes; (3) Evaluate the performance of hadronic showers with the crystal ECAL and HCAL options; (4) Study how to utilize the timing information of showers for the particle identification, shower separation, response compensation, etc.; (4) Explore potential of detecting both Cherenkov and scintillation photons with a crystal-SiPM detector unit and study the performance with the dual-gated or dual-readout schemes.

\subsubsection{RADiCAL -- Ultra-compact, Radiation-hard, Precision-timing Electromagnetic Calorimetry}

RADiCAL (Snowmass2021 LoI~\cite{LoI:Radical}) is a collaborative conducting R\&D on advanced calorimetry techniques for FCC-hh experimental applications and based on radiation-hard scintillator, waveshifter and photosensor technologies~\cite{RADiCAL}.   The emphasis is on ultra-compact, radiation hard EM calorimeters, based on modular structures consisting of alternating layers of very dense absorber and scintillating plates, read out via wavelength shifting solid fiber or capillary elements to photosensors positioned either proximately or remotely, depending upon their radiation tolerance. Initial versions of  RADiCAL modules consist of interleaved layers of 1.5mm thick LYSO:Ce tiles and 2.5mm thick tungsten plates, stacked to a total depth of $114\,{\rm mm}$, corresponding to $25\,{\rm X}_0$ and $0.9\,\lambda_I$.  The modular cross section of $14 \times 14\,{\rm mm}^2$, set by the $13.7\,{\rm mm}$ Molière radius of the structure, results in an ultra-compact module both transversely and longitudinally.   The shower energy is measured from the overall optical signal detected in the module; timing and position are measured from the optical signals detected in the region of EM shower max.  R\&D on rad hard materials is central to the effort and includes: for scintillators:  LYSO:Ce crystal plates, LuAG:Ce and LuAG:Pr ceramic plates and ${\rm BaF}_2$:Y crystal plates; for waveshifters: DSB1 organic filaments, LuAG:Ce ceramic filaments, flavenol organic filaments, and quantum dot siloxane filaments; for photosensors: rad-hardened SiPM, and large band-gap sensors based on GaInP, SiC, diamond and other materials. The design goals for the modules include the simultaneous measurement of shower energy, with resolution $\sigma_E/E = 10\%/\sqrt{E} \oplus 0.3/E \oplus 0.7\%$~\cite{Aleksa:2019pvl}, and shower timing, with $\sigma_t < 50\,{\rm ps}$, and shower position determined to within a few mm.

\subsubsection{High-rate Forward Calo (Direct Detection in PMTs)}
A new idea for calorimetry which will survive in the forward region of future colliders, high intensity experiments, and orbiting systems is the use of Photomultipliers (PMTs) for the direct detection of shower particles  (Snowmass2021 LoI~\cite{LoI:Forward}) . The PMTs act as direct calorimeter sensors to detect shower particles via Cerenkov light in the PMT window, and/or by direct secondary emission from shower particles traversing the dynodes. The secondary emission proportional to dE/dx provides compensating information.

Calorimeters in the Forward Region ($\eta \geq 3$) at future colliders will need to measure in a very hostile environment at irradiation levels $\sim 1\,{\rm GigaGy}$ and $>10^{17}$ neutrons/${\rm cm}^2$, with an expected number of pileup events of $100-1000$ per bunch crossing.  Almost none of the presently used calorimetry basic sensors can either survive the radiation damage or operate at the required rate or both. It has been proposed to study the capability of photomultiplier and dynode technologies as a potential energy-flow calorimeter with multiple signal compensation. Previous experience with dynodes in PMTs and with detectors in space and x-ray scanners show that metal-oxides can survive many GRad exposures. Depending on the application, PMTs could be manufactured with a metal “window” – no photocathode – for dynode secondary emission as the shower particles sampling detector. A PMT with a quartz or sapphire window and coupled to a quartz, sapphire,${\rm MgF}_2$ or silica aerogel tile may prove to be quasi-compensating, by in effect dual sampling. The Cerenkov light is more sensitive to EM showers, while the dynode secondary emission is proportional to dE/dx for shower particles traversing the PMT. Examination of the waveform may indicate differences in the Cerenkov signal and the direct dynode signal in the shower.

\subsubsection{Compact Readout Electronics}
Compact high-performance readout techniques are essential for a successful operation of particle flow calorimeters. Coordinated developments foster the common data taking with different prototypes and facilitate the comparison of the performance of different proposals. 

Within the CALICE collaboration, R\&D on a compact, homogeneous readout system for all the prototypes is ongoing (Snowmass2021 LoI~\cite{LoI:CALICEreadout}).
The most compact implementation so far is the so-called SL-Board for the CALICE silicon tungsten electromagnetic calorimeter (SiW ECAL), which has a length between 10 and 42mm and a width of 180mm and is conceived to serve up to 10k calorimeter channels of a detector layer. The cards are connected to a data concentrator system, able to serve up to 30 layers. It is envisaged to adapt the system of the SiW ECAL to other calorimeters. This task is facilitated by the fact that all calorimeters use a similar type of front-end ASIC. On the other hand electromagnetic and hadronic calorimeters differ in size and layout. Moreover additional functionality needs to be taken into account. The analogue hadronic calorimeter and the scintillator electromagnetic calorimeter for example use a LED system for the calibration of the scintillating tiles and the SiPMs, which has to be driven by the interface card. Detailed design work is still needed to understand how to transfer the concept in place for the SiW ECAL to for example hadron calorimeters. A major difference is the typical size of the prototypes and final systems which may require modifications in the way clock and control signals have to be propagated within the individual systems.

The heat dissipation of the SL-Board will require the integration of cooling. The SL-Board is designed for the integration of a cooling system, and cooling systems do exist, but detailed integration work has to start in the near future. 
In the coming years it is planned to investigate precision timing in highly granular calorimeters. The digital readout systems will have to cope with new requirements while at the same time keep the same level of compactness.

\subsection{Design Goals}
Since all calorimeter designs optimized for particle flow reconstruction algorithms feature small cells and a high channel count, they face a number of common challenges. On the other hand, their (intended) use in different environments as well as the characteristics of the employed detection techniques lead to design differences.

\subsubsection{Commonalities}
The cell size of particle flow calorimeters is typically influenced by several requirements:
\begin{itemize}
    \item in the ECAL, the cell should be small enough to allow the separation of close-by showers and a reliable assignment of showers produced by charged particles to the corresponding track in the tracking detector.
    \item in order to distinguish electromagnetic and hadronic showers as well as to identify electromagnetic sub-showers within hadron showers, the cell size should be of the order of the Moliere radius and the radiation length $X_0$, which are the typical scales of electromagnetic (sub-)showers
    \item a very small cell size especially in the ECAL can be beneficial for the reconstruction of $\tau$ decays 
    \item especially at hadron colliders, a small cell size allows a better rejection of energy depositions from pile-up events.
\end{itemize}
Typical cell sizes are in the range of $0.5 - 1.0\,{\rm cm}^2$ for ECALs  and $1.0$ to $\sim 10\,{\rm cm}^2$ for HCALs. A notable exception is the digital ECAL with a cell size of $30 \times 30\,{\rm \mu m}^2$.

The large channel densities mean it is not practical to bring the analog signals of all cells outside of the main calorimeter volume. In order to avoid this, the first step of the data acquisition including zero-suppression or some other means of data reduction is usually performed by front end electronics fully embedded in the active layers. This has several implications for the calorimeter design:
\begin{itemize}
    \item the fully embedded electronics is not accessible for repair, so has strong requirements on quality tests before assembly and on longevity
    \item complex readout elements such as ASICs will be situated in positions of vulnerability to radiation and environmental issues 
    \item the amount of heat produced by the electronics needs to be minimized or removed by cooling inside the active layers
    \item the large area covered by PCB on the active area can contribute significantly to the cost especially for HCALs.
\end{itemize}

\subsubsection{Distinguishing characteristics}

At future hadron colliders like HL-LHC and FCC-hh, the expected radiation levels restricts the choice of the active material in the most affected areas. Up to now, the most common materials are liquid Argon (or more general liquid noble gases) and silicon. For the latter, cooling to $-30^\circ {\rm C}$ or colder helps to keep the noise acceptable, but poses other practical problems.

For most PF calorimeter concepts, a "classical" analog readout is foreseen. The exceptions are the digital silicon ECAL and the (semi-)digital gaseous HCALs. For the ECAL, the availability of Monolithic Active Pixel Sensors (MAPS), which were originally developed for vertex and tracking detectors, lead to the idea of a calorimeter with very small cell size and digital readout, which can provide unrivaled position resolution and two-particle separation. For the energy resolution, the small sampling fraction of thin silicon sensors as well as saturation effects at high energies, when more than one shower particle hits a pixel, play a role. 
For gaseous HCALs with detectors operated in avalanche mode (e.g. RPCs), the measured hit energy is strongly influenced by the depth at which the avalanche started, and therefore is not really proportional to the energy deposition of the traversing shower particles. This makes a simplification of the readout system by reading out only digital or semi-digital hit information attractive. 

The various materials and technologies discussed above have some characteristics that lead to differences in the detector design:
\begin{description}
\item[Silicon:] Silicon sensors show very stable signals with very small channel-to-channel variations, but they are expensive, and the signals are very small, which means that special care has to be taken in the design of the readout system.

\item[Crystals and Ceramics:] The high density, radiation hardness and intrinsic brightness of scintillation crystals and inorganic ceramic scintillator plates make them attractive choices in particular for the EM region of PF calorimetry.  Detector designs include their use as active layers in sampling calorimetry and optically transparent blocks in homogeneous designs.  A strength of the approach is the prospect of good sampling fraction, to the  benefit of both energy and timing resolution.   A challenge is the cost of scintillation crystal manufacture, which has helped to drive the development of more cost-effective alternatives based on ceramic scintillators. 

\item[Gaseous:] The signal of gaseous detectors like RPCs or MPGDs is  sensitive to the gas flow and to environmental conditions like temperature and humidity. This means that special attention has to be paid to the monitoring of these conditions and of the stability of the detector response. Many of the common gas mixtures used in gaseous detectors have a strong negative impact on the environment, so either the detector has to be built leak-tight or R\&D on eco-friendly gas mixtures is mandatory.
\item[SiPM-on-tile:] The SiPM signal for a given number of photons depends on the overvoltage, the difference between the bias voltage and the breakdown voltage. Since the breakdown voltage is sensitive to the temperature, either the temperature needs to be kept stable or the bias voltage should be adjusted to keep overvoltage fixed. At large hit amplitudes, saturation effects due to the limited number of pixels play a role, which can be corrected at analysis level if they are known well enough.
\end{description}
Different active materials can also lead to different mechanical considerations. Examples are the fragility of silicon sensors or the rigidity of glass RPCs, which need very flat absorber plates and very well defined gaps in the absorber.

\subsection{How they are built}
\subsubsection{Structure and Sensors}

An overview of the characteristics of the calorimeter concepts discussed in the previous sections can be found in Table~\ref{tab:overview}.

\begin{table}[htb]
\renewcommand{\arraystretch}{1.3} 
\centering
{\scriptsize
\begin{minipage}{\linewidth}
\begin{tabular}{|l|l|l|l|l|l|l|}
\hline
  name   &  purpose & project & active & channel size & readout  & $\#$ of layers  \\ 
   & & & material &   & (A/D/SD) & (depth)  \\ \hline
  CALICE SiW ECAL  & ECAL & ILC~\footnote{also for CLIC \& FCC-ee} & silicon & $5 \times 5\,{\rm mm}^2$ & analog & $30$ ($24 X_0$)   \\
  SiD ECAL   & ECAL & ILC & silicon & $13\,{\rm mm}^2$ & analog & $30$ ($26 X_0$) \\
  HGCAL Si & ECAL~\footnote{silicon also used in HCAL part} & HL-LHC & silicon & $52 - 118\,{\rm mm}^2$ & analog & $28$ ($25 X_0$) \\
  FoCal &  ECAL & HL-LHC & silicon & $30 \times 30\,{\rm \mu m}^2 $ & digital & $28$ ($25 X_0$) \\
  CALICE Sci-ECAL & ECAL & ILC~\footnote{also for CEPC} & SiPM-on-tile & $5 \times 5\,{\rm mm}^2$~\footnote{effective size, strips have $5 \times 45\,{\rm mm}^2$} & analog & $30$ ($24 X_0$)  \\
  RADiCAL & ECAL & FCC-hh & crystal + WLS &  $4 \times 4\,{\rm mm}^2$~\footnote{effective shower size at shower max; module cross section is  $14 \times 14\,{\rm mm}^2$} & analog & $29$ ($25 X_0$)    \\ \hline
  CALICE AHCAL & HCAL & ILC~\footnote{also for CEPC, CLIC \& FCC-ee} & SiPM-on-tile & $3 \times 3\,{\rm cm}^2$ & analog & $40$ ($4 \lambda_I$) \\
  HGCAL Scint & HCAL & HL-LHC & SiPM-on-tile & $6 - 30\,{\rm cm}^2$ & analog & $22$ ($7.8 \lambda_I$)~\footnote{contains also pure silicon and mixed layers}  \\
  CALICE DHCAL & HCAL & ILC & RPC & $1 \times 1\,{\rm cm}^2$ & digital & $40$ ($4 \lambda_I$)  \\
  CALICE SDHCAL & HCAL & ILC & RPC & $1 \times 1\,{\rm cm}^2$ & semi-digital & $40$ ($4 \lambda_I$)   \\
  \hline 
\end{tabular}
\end{minipage}
}
\caption{Overview of the characteristics of several particle flow calorimeter concepts and technologies. 
}
\label{tab:overview}
\end{table}

In addition to these calorimeter concepts, new active materials have been proposed for particle flow calorimeters, for which quantities like channel size and depth are not yet defined and could not be included in the table. These comprise highly granular scintillator  crystals, advanced MPGDs like Micromegas and classical photomultipliers used directly for the detection of shower particles.

\subsubsection{Fabrication, Assembly, Factory-style methodology.}
Due to the large channel count, scalable production and assembly technologies are important for all PF calorimeter concepts. In order to reduce manual assembly steps and the number of units that have to be handled manually, large basic units, called modules in the following, are attractive. Some examples are:
\begin{description}
\item[Silicon:] For silicon ECALs, the module size is related to the wafer size from which the sensors are produced. In the case of the CALICE SiW ECAL, the sensors have a square shape and are made from 6-inch wafers. The module is made from one PCB equipped with 4 silicon sensors. For the CMS HGCAL, hexagonal sensors made from 8-inch wafers have been chosen because they use a larger fraction of the wafer area for the sensor. The module comprises one sensor. For both calorimeters (semi-)automatic module assembly stations for the gluing of the sensor(s) to the PCB, and in the case of the HGCAL for the wire bonding, have been built. 
\item[RPC:] In the CALICE SDHCAL prototype, the individual chambers have a size of $1\,{\rm m}^2$. For a detector at ILC, individual chambers would cover several square meters. These sizes are too large to be covered with a single PCB, so the basic unit size is given by the PCB size, which is determined by the size that commercial PCB manufacturers and assembly companies can handle. 
\item[SiPM-on-tile:] The module assembly, which comprises the assembly of electronics components including SiPMs as well as the gluing of individual scintillator tiles or strips, can be performed with commercial pick-n-place machines. This has been demonstrated with the CALICE AHCAL prototype, and is planned to be used for the CMS HGCAL. An alternative concept are {\it megatiles}, where the full module is covered by one large scintillator. The optical separation between neighboring channels is obtained by milled grooves filled with reflective material. In both cases, the SiPM-on-tile module size is determined by a practical PCB size.
\end{description}

\subsubsection{Industrialization}
Even though the overall channel counts of PF calorimeters are hundred-thousands to millions and the covered areas in the active layers are hundreds of square meters, these products are a niche for most industry areas. Usually, only the components and some of the first assembly steps can be bought from or done by companies. Examples are silicon sensors (but the requirements are unusual, so there are very few suppliers), SiPMs, Micromegas, and production and component assembly of PCBs. The production of modules (joining the active sensor with the readout electronics) and everything from there on is typically not possible in industry.

\subsection{Power Implications}
As mentioned above, the large channel count of PF calorimeters usually implies having the front end electronics fully embedded in the active layers. The amount of heat produced by the electronics needs to be minimized or removed by cooling inside the active layers.
\subsubsection{$\mathbf{e^+e^-}$ Linear Collider – Power Reduction Between Interactions}
 Detectors planned for linear electron-positron colliders like ILC and CLIC can use the bunch train structure of the beam to minimize the power consumption. In these calorimeter concepts, no active cooling is foreseen, in order to avoid dead regions in the detector and keep the material as homogeneous as possible. The readout electronics is adapted to the low duty cycle of a linear electron-positron collider, in which short bunch trains ($1\,{\rm ms}$ long for ILC) are interleaved with long gaps without particles ($199\,{\rm ms}$ long for ILC). This is taken into account in the readout electronics by having two phases: a data acquisition phase in which the ASICs are acquiring and storing the data, and a readout phase in which the data are digitized and sent off detector. During the readout phase, all non-necessary components of the ASIC can be switched off {\it (power pulsing)}.
 Typical numbers for the power consumption goal are tens of ${\rm \mu W}$ per channel. The consumption in power pulsing mode compared to continuous running is reduced down to a few percent.
 This running mode is not possible for colliders without a similar bunch train structure, so it can be applied neither at circular electron-positron nor at hadron colliders.
 
\subsubsection{Hadron Colliders - Systemic Heating and Cooling}
For colliders where the beam structure does not allow the use of power pulsing, the heat has to be removed by the cooling system. As an example, the power consumption of the CMS HGCAL is expected to be $\sim 110\,{\rm kW}$ per endcap, of which roughly 2/3 comes from the readout electronics~\cite{CERN-LHCC-2017-023}. The power consumption of the sensors is less than $10\%$ after the radiation damage accumulated during an integrated luminosity of $3000\,{\rm fb}^{-1}$. The HGCAL foresees a dedicated ${\rm CO}_2$ cooling system with the modules in the active layers mounted on copper cooling plates. A similar power consumption would also be expected for a detector at a circular electron-positron collider. At hadron colliders, the mitigation of the effects of radiation damage might pose additional requirements on the operation conditions. For the HGCAL, the operating temperature is foreseen to be $-30^\circ\,{\rm C}$ to reduce the radiation-induced noise in the silicon sensors as well as in the SiPMs. 

\subsection{Cost Implications}
Due to the large channel count, readout electronics are an important cost driver for PF calorimeters. As example, for the CMS HGCAL the electronics and electrical system together with the back-end (trigger and DAQ) system make up nearly one third of the total cost, and are the second largest cost item after the silicon sensors~\cite{CERN-LHCC-2017-023}. For hadron calorimeters with their large areas of active layers, the cost of the PCB production and assembly can be significant, independent of the active material that is used. 


\section{Event Reconstruction Techniques}
In general, the high granularity and large channel count are not only a challenge for the design and construction of a particle flow calorimeter, but also for the software needed to simulate and analyze the data.
\subsection{Simulation Methodologies} 
Most detector simulation packages rely on the Geant4 toolkit~\cite{ref:Geant1,ref:Geant2,ref:Geant3} for the simulation of the passage of particles through matter. As discussed in Section~\ref{section:performance}, in the Geant4 versions available in recent years, simulations with the recommended physics lists agree with available testbeam data with an accuracy of a few percent for most choices of absorber and sensor material. Therefore, the simulations can be used to optimize the designs. 

Special attention is needed for calorimeters with gaseous active material, where the agreement with data is typically less good than for other active materials. In addition, for these detectors the digitization, which describes the step from the energy deposition by a traversing particle in the active material to the recorded hit (or hits) in the detector, needs to be handled carefully. For gaseous detectors operated in avalanche mode, the size of the avalanche and the recorded energy depend on the depth within the active layers where the avalalche starts, which needs to be taken into account in the digitization. 

For detectors with neutron-rich absorber material (e.g. tungsten) and neutron-sensitive detection technique (e.g. plastic scintillator), it is important that low-energy neutrons are tracked with enough precision. Geant4 provides a set of physics lists labeled "high precision" (with suffix \_HP in the name). This is essential for a good agreement of the simulation with data.

The large number and small volume of readout cells (e.g. silicon pads or scintillator tiles/strips) usually makes handling of these as individual detection volumes impractical in Geant4. However, simulating larger volumes and subdividing these in the digitization by particle impact point or the position of the energy deposition step in Geant4 works well.

Even with tricks like this, the simulation of PF calorimeters can be very time consuming. Machine Learning techniques are being developed to speed up the process by training the models either with samples simulated with Geant4 as input, or with measurements from testbeam prototypes. First results look encouraging, but are not yet at a stage where they can (partly) replace Geant4 in the production of general purpose simulation samples.

\subsection{Calibration Methodologies}

The absolute calibration of individual cells in a PF calorimeter is less important than in "classical" calorimeters, where the energy of a particle shower is contained in a few cells, while typically hundreds of cells contribute to the measurement of a shower energy in highly granular calorimeters. Therefore, many small calibration errors in the individual cells can cancel out, meaning that they will not lead to a bias in the reconstructed energy, but can have an influence on the resolution. For hadron calorimeters, usually the intrinsic fluctuations of the hadron showers are so large that the influence of the calibration errors on the resolution is negligible. These individual cell errors can only cancel and lead to no bias, if they are statistically distributed. If all cells have an error in the same direction, or if there are large regions with error in the same direction, this is no longer true. Therefore, for the large number of cells, efficient calibration methods are needed that avoid these possible biases. They also need to allow the long-term monitoring of the detector stability. In addition to dedicated calibration systems, the signal of minimum-ionizing particles like muons can be used as reference in the calibration: either the most probable hit amplitude, or (if no amplitude information is recorded) the efficiency and hit multiplicity. 

\subsection{Analysis Methodologies}
The measurement of jets with particle flow algorithms in detectors optimized for this reconstruction technique has originally been proposed for linear electron-positron Higgs factories~\cite{Brient:2001fow,Morgunov:2002pe}. Since then, several software frameworks for particle flow reconstruction have been developed. For Higgs factories, the Pandora PFA~\cite{Thomson:2009rp} is widely used and performing very well. Pandora uses both the topological information and the energy measurement in the calorimeter to assign calorimeter cells with energy deposits to clusters and calorimeter clusters to tracks. The Pandora Software Development Kit~\cite{Marshall:2015rfa} aims to provide a robust, reliable and easy-to-use environment for developing and running pattern recognition algorithms. An alternative PFA, ARBOR~\cite{Ruan:2013rkk}, is based only on the topological information. Recently, ideas from deep learning and computer vision~\cite{DiBello:2020bas} have been proposed for improved PF reconstruction.

PFAs are also used and tuned for detectors which were not optimized for their application. Current examples are the PFA algorithms employed by the ATLAS~\cite{ATLAS:2017ghe} and CMS~\cite{CMS:2017yfk} experiments. In the harsh environment at the LHC, it is essential for a reliable reconstruction of jets and missing energy to apply pile-up mitigation algorithms in addition.

\section{Assessing Performance of Current PF Calorimeters}
\label{section:performance}
Several of the design goals of particle flow calorimeters depend on the performance of the whole detector system including the tracker, and on the response to many impacting particles, e.g. for the jet energy resolution reconstructed with a PFA or for the suppression of pile-up. Therefore, it is difficult or not possible to assess these quantities in beam tests, where usually only limited information from tracking detectors and only single particles are available. Nevertheless, beam tests provide important insight into some basic performance measures of the calorimeters, and into the agreement with simulation programs that are used to determine the expected performance of the complete detector system with real data. Long term stability of the detector technologies can be evaluated to some extent by beam tests with the same prototype over several years, although this is not directly comparable to the continuous operation over several years in a collider experiment.

\subsection{Results From Beam Testing}
An overview of the results of PF calorimeters in beam tests until 2015, mainly conducted within the CALICE collaboration, can be found in~\cite{Sefkow:2015hna}. Since then, progress has been made within CALICE on further beam test analyses of existing prototypes as well as construction and first beam tests of technological prototypes of the AHCAL and the SiW ECAL. For the SiD ECAL, results have been obtained from an initial 9-layer prototype and further R\&D is foreseen on a full-depth stack. For the CMS HGCAL, several beam tests have been performed ranging from single-layer setups to a system consisting of an electromagnetic and a hadronic silicon section and a SiPM-on-tile section.

For PF Calorimeter designs, notably those instrumented with silicon detectors to provide high spatial resolution such as the CMS HGCAL, the attendant sampling fraction is very small (few \%) impacting EM energy resolution.  This is not an oversight, but rather represents a design choice taken by the collaboration.

\subsubsection{CALICE}
Within CALICE, all technologies (SiW ECAL, Sci ECAL, AHCAL, DHCAL and SDHCAL) have been studied with large prototypes in muon, electron and hadron beams, either with a physics prototype or a technological prototype. Also combined tests have been performedof an ECAL and an HCAL to validate the performance of the calorimeter system. 

The strategy in all cases is to use the response to mips (muons) to monitor the stability of the detector and to derive a calibration at low hit energies to: assess the response to electrons or positrons with their well-known shower development to validate the detector modeling of the prototypes and to study the behavior at high hit energies, and assess the response to hadrons to study how well hadronic showers are described by the simulation. In addition, single-particle reconstruction algorithms including software compensation and important ingredients into particle flow algorithms like the separation of close-by showers have been investigated.

The most important findings in~\cite{Sefkow:2015hna} were: "[...] all technologies have been successfully commissioned, have recorded large data sets and, in the results presented so far, demonstrated their basic performance. No show stoppers were found. Therefore the proof of principle based on the full analyses holds for the feasibility of particle-flow type reconstruction in fine-grained calorimeters in general. [...] The CALICE data confirm the progress made in recent years in improving the theory-motivated shower models implemented in the Geant4 framework, which had been driven by test beam studies with LHC calorimeter prototypes [...] Overall, the more recent hadron shower models reproduce the pion and proton data in the relevant energy range with an accuracy of a few percent. Shower shapes are well described in terms of average depth, while their width remains somewhat under-estimated as energies increase. [...] A test of the particle flow reconstruction methods was performed by applying the Pandora algorithm to test beam events overlaid on each other, and quantifying confusion effects by measuring the degradation of neutral hadron energy measurement through the presence of close-by charged hadron showers, and comparing results with real data to those with simulations. [...] The excellent agreement represents important underlying evidence for the applicability of the method for jet measurements at future colliders."

Since then, further studies of electron showers measured with the DHCAL physics prototype (without absorber)~\cite{CALICE:2016qwo} and hadron showers with the SiW ECAL physics prototype~\cite{CALICE:2019vza}, the AHCAL physics prototype with steel~\cite{CALICE:2016nds} and tungsten absorber~\cite{CALICE:2015fpv}, the DHCAL physics prototype~\cite{CALICE:2019rct} and the SDHCAL prototype~\cite{CALICE:2016fbz,CALICE:2017nol} have been published. Overall, they confirm the assessments made in 2015. 

For several technologies developed within CALICE, technological prototypes have been constructed or are being constructed.
The large AHCAL technological prototype has been tested with particle beams in 2018, and a paper on the construction and commissioning is in preparation. Further publications on shower analyses as well as further beam tests are planned. For the SiW ECAL and the SiECAL, the construction of the technological prototypes is in progress (or completed), but no beam test results of multi-layer systems are available as yet.

\subsubsection{FoCal} 
In the context of the R\&D for the ALICE FoCal upgrade, a prototype of a Si-W EM calorimeter was built with Monolithic Active Pixel Sensors as the active elements~\cite{deHaas:2017fkf}. It was based on MIMOSA chips with a pixel size of $30\,{\rm \mu m}$ and had 39 million pixels in 24 layers. Energy and position resolution for electron showers were determined, and the shape of the showers was measured in unprecedented detail.

\subsubsection{SiD}
Recent results obtained with the 9-layer prototype in 12.1 GeV electron beam have been shown at conferences~\cite{Steinhebel:2017qze}, and are in reasonable agreement with simulation. The testing did uncover an important issue related to many pixels being triggered simultaneously.  Such systemic issues are not unique to SiD, and are important and challenging in PF designs, necessitating the need for extensive system testing.

\subsubsection{CMS HGCAL}
During the R\&D for the CMS HGCAL, beam test of prototype modules in various configurations have been performed.
First hexagonal silicon modules, using the existing Skiroc2 front-end ASIC developed for CALICE, have been tested in beams at Fermilab and CERN in 2016~\cite{Akchurin:2018rpm}. With up to 16 modules, system stability, calibration with minimum-ionizing particles and resolution (energy, position and timing) for electrons were determined, and were found to be described well by the GEANT4-based simulation. 

Next, a larger prototype was built based on modules constructed with 6-inch hexagonal silicon sensors with cell areas of $1.1\,{\rm cm}^2$, and the SKIROC2-CMS readout ASIC ~\cite{CMSHGCAL:2020dnm}. Beam tests of different sampling configurations were conducted with the prototype modules at DESY and CERN in 2017 and 2018. The up to $\sim 12,000$ channels of silicon sensors were arranged in up to 28 layers of single modules mounted on copper cooling plates for the electromagnetic section and up to 12 layers with up to 7 neighboring modules for the hadronic section. Measurements of the prototype energy resolution and linearity, position resolution, resolution on the positron angle of incidence derived from the shower axis reconstruction and shower shapes were performed and compared to detailed GEANT4 simulations~\cite{CMSHGCAL:2021nyx}. For the measurement of hadron showers, the silicon section was complemented with the CALICE AHCAL technological prototype as SiPM-on-tile hadronic calorimeter. Publications on the hadron shower measurements as well as the hit time measurement are in preparation.

The HGCAL prototype modules used in the beam tests differ in several aspects from the final HGCAL design, e.g. in the size of the silicon sensors and in the readout electronics, so the final assessment of the HGCAL performance will only be possible with further beam tests of (pre-)production modules or the HGCAL itself. 

\subsection{Results from experimental operation} 
To date there has been no direct experimental application of particle flow calorimeters, only results from extensive beam testing. The first full demonstration of the technique will be the CMS HGCAL, which will be the first to break ground in this new technological territory.

\section{Future Directions of the Particle Flow Technique}
The Particle Flow approach has emphasized performance enhancement through the effective extension of particle tracking into the calorimetry itself, facilitated by very high spatial granularity.  This is an appropriate match to ILC operation, where few interactions are expected during beam collisions.  However at circular machines, and particularly hadron colliders, event pileup and radiation backgrounds are a fundamental challenge, as are the association of identified particles with the multitude of possible production vertices created during beam crossings. Hence precision timing of particles is also significant, and is important to the detector designs.  New technologies continue to be developed, some presented in this summer study, which directly address the simultaneous, precision measurement of energy, position and time of particles, furthering the development of Particle Flow Calorimetry.

\section{Assessment of long-term importance of PF Calorimeters to the field}
    The promise of Particle Flow Calorimetry is its potential for excellent jet resolution and identification of neutral hadron energy, performance measures that are challenging to reach through other calorimetric approaches.  While initially envisaged and designed to address the challenging requirements of Higgs Factories such as the ILC, the first actual experimental application will be provided in the hadron collider environment at  the HL-LHC, with the CMS HGCAL. When it is brought into operation later in this decade, the HGCAL  will serve as a "stress test" of the PF approach, and will provide important information about future application of such instrumentation at hadron colliders such as FCChh and circular lepton colliders such as FCCee, where power considerations, event pileup or radiation backgrounds are significantly different than for linear lepton colliders.  The impact of the Particle Flow Calorimetry Technique rests on demonstrations of its successful experimental application.  The CMS HGCAL will be the first of these.
\clearpage

\bibliographystyle{JHEP}

\bibliography{bibliography}


\end{document}